# How do Globally Distributed Agile Teams Self-organise?
## *Initial Insights from a Case Study*

Sherlock A. Licorish and Stephen G. MacDonell
*SERL, School of Computing & Mathematical Sciences, Auckland University of Technology,
Private Bag 92006, Auckland 1142, New Zealand*
*{sherlock.licorish, stephen.macdonell}@aut.ac.nz*

**Abstract**

*Agile software developers are required to self-organize, occupying various informal roles as needed in order to successfully deliver software features. However, previous research has reported conflicting evidence about the way teams actually undertake this activity. The ability to self-organize is particularly necessary for software development in globally distributed environments, where distance has been shown to exacerbate human-centric issues. Understanding the way successful teams self-organise should inform distributed team composition strategies and software project governance. We have used psycholinguistics to study the way IBM Rational Jazz practitioners enacted various roles, expressed attitudes and shared competencies to successfully self-organize in their global projects. Among our findings, we uncovered that practitioners enacted various roles depending on their teams' cohort of features; and that team leaders were most critical to IBM Jazz teams' self-organisation. We discuss these findings and highlight their implications for software project governance.*

## 1. INTRODUCTION

The consensus of recent evidence continues to indicate that a variety of human and social factors are among the strongest determinants of software development project performance (e.g., see (Abrahamsson et al., 2006)). In particular, the matching of software practitioners to certain roles has been shown to aid with task performance (Acuna et al., 2006), lending credence to an assumption that particular software activities demand specific competencies, and individuals who possess higher levels of those competencies would perform best in the corresponding roles. The implication of such a finding is that role assignment should be conducted in relation to individuals' specific expertise.

Proponents of Agile software development have challenged these views however, and methods such as Extreme Programming (XP), Adaptive Software Development (ASD), and SCRUM all emphasise the need for self-organisation and less rigid team role assignment (Pressman, 2009). Additionally, studies examining self-organizing agile teams have found evidence that suggests that project team members do indeed adopt various roles, as needed, to facilitate self-organization during projects. Hoda et al. (2010), for instance, found that the roles of mentor, translator, champion, coordinator, promoter and terminator were assumed at various times by different team members so that project management, team harmony and effective idea generation were sustained during development.

However, while the adoption of roles in such a way is said to be evident and necessary in agile development contexts (Hoda et al., 2010), previous work has noted that it is rarely achieved (Moe et al., 2008). In addition, investigations of the issues of expertise, role assignment, role adoption and self-organization in globally distributed development contexts have not been reported. This is despite the relevance of these phenomena in such settings, given the often limited opportunities available for rapid communication and feedback (Serce et al., 2009). We have therefore used psycholinguistics to analyze repository data to examine the specific attitudes and competencies adopted by those occupying a range of roles while they were working in multiple teams during distributed agile software development. Through this preliminary study we provide explanations for the way agile teams actually self-organize, along with recommendations for agile team composition and project governance.

In Section 2 we survey related work and outline our research questions, and our research settings are outlined in Section 3. Section 4 provides our measures, and our results and analysis are presented in Section 5. Section 6 provides a discussion of our main findings, highlights the study's implications, and outlines our future work. In Section 7 we consider our study's limitations.

## 2. RELATED WORK

Roles reflect the particular rights, responsibilities, expectations and behaviours that persons are expected to honour or fulfil (Belbin, 2002). The idea of studying and relating participants' behaviours to roles has attracted extensive research in the psychology, sociology and management disciplines (Ashforth, 2001; Biddle and Thomas, 1966; Hellriegel and Slocum, 2007). Research in these domains seeks to inform the process of personnel assignment to jobs or tasks according to their traits, based on the expectation that team role principles are relevant for informing effective team formation.

Outside of these domains, human resource management has also integrated psychology and role theories to support the task of selecting individuals with appropriate skill-sets for particular positions. In particular, most software-related positions demand multiple competencies, including intrapersonal, organisational, interpersonal and management skills (Downey, 2009). Intrapersonal skills include judgement, innovation and creativity, and tenacity, while having knowledge of specific environments (for example: recent programming competence in Java) is characterised as organisational. Inter-personal skills comprise team work, cooperation and negotiating skills, and management skills are related to planning, organisation and leadership.

In relation to software groups or departments, roles may also relate to the specific software process or methodology being utilised by teams. For instance, a software department adopting XP may define roles such as programmer, tester, coach and so on (Highsmith, 2004). Additionally, sometimes roles may be performed arbitrarily by project members in which case these environments require that team members possess general competency in many roles (Gorla and Lam, 2004). Thus, role arrangement and competency requirements for individual software-related roles are somewhat subject to specific organisational requirements and contexts (Trigo et al., 2010).

Psychology and role theories have also been applied previously in the software engineering discipline with success (Licorish et al., 2009 ). Such considerations were embedded in the view that human involvement, and the constraints that arise as a result of human issues in software development, substantially determine the outcomes of software projects (Standish Group, 2009). To this end, it has been asserted that studying issues related to how software developers behave in teams while solving problems may provide valuable insights into software development critical success factors.

As noted in Section 1, software engineering research has also considered teams from a self-organisation perspective. In fact, the ability to self-organize is purported to be one of the determinants of agile teams' success (Hoda et al., 2010). However, Moe et al. (2008) noted that the actual self-organising process is quite complex. Their ethnographic study in Norway of software developers utilizing agile practices uncovered that team members displayed little internal autonomy and were rarely willing to assume roles other than those that matched their specialized competencies. These findings are somewhat in contradiction to the work of Hoda et al. (2010), which found that agile developers in India and New Zealand operated more fluidly across informal roles.

The divergence in findings between the studies of Moe et al. (2008) and Hoda et al. (2010) suggests a need for additional research, to further examine the ways agile teams actually self-organise. In fact, the teams studied by Moe et al. (2008) were composed of novice developers, a factor that could have influenced the effects observed by these authors. This adds support for the view that further studies are required to address this issue, to provide understandings of how different attitudes and competencies are actually enacted by those assigned to specific roles during successful agile projects. As noted above, the ability to work across roles is particularly relevant for distributed agile software development settings, where issues related to distance, culture and personality demand that such teams are effective at self-organizing (Serce et al., 2009). We look to address this research opportunity by answering the following research questions:

*RQ1. How are roles enacted during distributed agile software development?*

*RQ2. Are the specific attitudes and competencies enacted related to practitioners' actual role assignment?*

## 3. RESEARCH SETTING

To address the research questions just specified we conducted a case study, and have examined artefacts and messages extracted from a specific release (1.0.1) of Jazz (based on the IBM[R] Rational[R] Team Concert[TM] (RTC) – refer to the Acknowledgements section for details). Jazz, created by IBM, is a fully functional environment for developing software and managing the entire software development process (Frost, 2007). The software includes features for work planning and traceability, software builds, code analysis, bug tracking and version control in one system, and so data captured in Jazz is likely to reflect what actually happens during the software development process. Changes to source code in the Jazz environment are only allowed as a consequence of work items (WIs) being created beforehand, in the form of a bug report, a new feature request or a request to enhance an existing feature. The Jazz repository comprised a large amount of process data from development and management activities carried out across the USA, Canada and Europe. Jazz teams use the "Eclipse-way" approach for guiding the software development process (Frost, 2007). This approach outlines iteration cycles that are six weeks in duration, comprising planning, development and stabilizing phases. Builds are executed after project iterations. All information for the software process is stored in a server repository that is accessible through a web-based or Eclipse-based RTC client interface.

While criteria for software project success are often said to relate to projects being completed on time, on budget and with the required features and functionality (Standish Group, 2009), others assert that measures related to software projects' impacts on the host organization, post-release customers' reviews and actual software usage are also relevant project success indicators (Espinosa et al., 2006). Accordingly, given the impact IBM Rational products (included in the Jazz repository, see jazz.net for details) have had on IBM and many other organizations (with over 30,000 companies using these tools), and that these products have been positively reviewed and tested by those companies, we would infer that Jazz teams should indeed be considered as successful. Thus, studying these teams should provide us with insights into successful teams' enacted roles, attitudes and competencies.

We created a Java program to leverage the Jazz Client API to extract information and development and communication artefacts from ten teams (shown in Table 1) from the Jazz repository. This included: Work Items (WIs) and history logs, representing project management and development tasks; Project Workspaces, representing multiple team areas and including information on team memberships and roles; and Messages, representing practitioner dialogues and communication around project WIs.

The selected project artefacts amounted to 1201 software development tasks, involving 394 contributors belonging to five different roles (described below), and 5563 messages exchanged around the 1201 tasks. As the data were analyzed, it became clear that the ten cases selected were representative of those in the repository, as we reached data saturation (Glaser and Strauss, 1967) after analyzing the third project case. Additionally, we used social network analysis (SNA) to explore the teams' communications and noted that all ten teams had similar profiles for network density (between 0.02 and 0.14) and closeness (between 0 and 0.06). Formal statistical testing for significant differences in In-Degree measures also confirmed that the projects were relatively homogenous, $X^2 = 13.182$, $p = 0.155$ (Kruskal-Wallis test result).

The actual role information extracted from the repository is as follows: Team leads (or component leads) are responsible for planning and executing the architectural integration of components; Admins are responsible for the configuration and integration of artefacts; Project managers (PMC) are responsible for project governance; those occupying the Programmer (contributor) role contribute code to features; and finally, those who occupied more than one of these roles were labelled Multiple. We used these practitioners' roles as our unit of analysis, we made comparisons of attitude and competencies across roles in individual teams, and we also conducted assessments across various task types.

## 4. MEASURING ENACTED ROLES, ATTITUDES AND COMPETENCIES

Previous research has identified that an individual's linguistic style is quite stable over time and that text analysis programs are able to accurately link language characteristics to attitudes (e.g., see (Mairesse et al., 2007)). We employed the Linguistic Inquiry and Word Count (LIWC) software tool in our analysis of practitioners' messages to discern the roles they enacted during their projects and the way specific attitudes and competencies were enacted given practitioners' actual roles. The LIWC is a software tool created after four decades of research using data collected across the USA, Canada and New Zealand (Pennebaker and King, 1999). Similar to an electronic parser, this tool accepts written text as input which is then processed based on the LIWC dictionary, after which summarized output is provided. The several linguistic dimensions assessed by the tool and reported in the summary (see Table 2) are said to capture the psychology of individuals through the words they use (Pennebaker and King, 1999, Mairesse et al., 2007). For example, consider the following sample comment:

> "We are aiming to have all the patches ready by the end of this release; this will provide us some space for the next one. Also, we are extremely confident that similar bug-issues will not appear in the future."

**Table 1**. Summary statistics for the selected Jazz projects.

| Project (Team) ID | Task (WI) Count | Software Tasks (Project) | Total Contributors – Roles | Total Messages | Period (days) – Iterations |
|---|---|---|---|---|---|
| P1 | 54 | User Experience – tasks related to UI development | 33 – 18 programmers, 11 team leads, 2 project managers, 1 admin, 1 multiple roles | 460 | 304 - 04 |
| P2 | 112 | User Experience – tasks related to UI development | 47 – 24 programmers, 14 team leads, 2 project managers, 1 admin, 6 multiple roles | 975 | 630 - 11 |
| P3 | 30 | Documentation – tasks related to Web portal documentation | 29 – 12 programmers, 10 team leads, 4 project managers, 1 admin, 2 multiple roles | 158 | 59 - 02 |
| P4 | 214 | Code (Functionality) – tasks related to development of application middleware | 39 – 20 programmers, 11 team leads, 2 project managers, 2 admins, 4 multiple roles | 883 | 539 - 06 |
| P5 | 122 | Code (Functionality) – tasks related to development of application middleware | 48 – 23 programmers, 14 team leads, 4 project managers, 1 admin, 6 multiple roles | 539 | 1014 - 17 |
| P6 | 111 | Code (Functionality) – tasks related to development of application middleware | 25 – 11 programmers, 9 team leads, 2 project managers, 3 multiple roles | 553 | 224 - 13 |
| P7 | 91 | Code (Functionality) – tasks related to development of application middleware | 16 – 6 programmers, 7 team leads, 1 project manager, 1 admin, 1 multiple roles | 489 | 360 - 11 |
| P8 | 210 | Project Management – tasks under the project managers' control | 90 – 29 programmers, 24 team leads, 6 project managers, 2 admins, 29 multiple roles | 612 | 660 - 16 |
| P9 | 50 | Code (Functionality) – tasks related to development of application middleware | 19 – 10 programmers, 3 team leads, 4 project managers, 2 multiple roles | 254 | 390 - 10 |
| P10 | 207 | Code (Functionality) – tasks related to development of application middleware | 48 – 22 programmers, 12 team leads, 2 project managers, 1 admin, 11 multiple roles | 640 | 520 - 11 |
| ∑ | 1201 | - | **394 contributors,** comprising 175 programmers, 115 team leads, 29 project managers, 10 admins, 65 multiple roles | **5563** | - |

In the comment the author is expressing optimism that the team will succeed, and in the process finish ahead of time and with acceptable quality standards. In this quotation, the words "we" and "us" are indicators of team or collective focus, "all", "extremely" and "confident" are associated with certainty, while the words "some" and "appear" are indicators of tentative processes. Words such as "bug-issues" and "patches" are not included in the LIWC dictionary and so would not affect the context of its use - whether it was to indicate a fault in software code or a problem with one's immunity to a disease. Although these omissions may be thought to represent a limitation of the approach, we know that the context is software development; and what is of interest, and is being captured by the tool, is evidence of attitudes and competencies. Previous work has also provided confirmation of the utility of the LIWC dictionary in a software development context (Rigby and Hassan, 2007). In the current work we examine practitioners' enacted roles, attitudes and competencies via their messages (5563 messages contributed by 394 practitioners assigned to five different roles), along multiple linguistic categories. Table 2 describes the categories chosen with brief theoretical justification for their selection.

## 5. RESULTS AND ANALYSIS

We grouped the 5563 individual messages based on practitioners' assigned project roles (Multiple, Team lead, Admin, Project manager or Programmer). We then examined the distributions of the five groups' use of linguistic dimensions for normality, using the Kolmogorov-Smirnov test. This revealed violations of the normality assumption. We therefore conducted Kruskal-Wallis tests to check for differences in the 16 linguistic dimensions (shown in Table 2) between those occupying the five roles. These tests revealed that there were statistically significant differences ($p < 0.05$) in language usage for each of the 16 linguistic dimensions between practitioners occupying the five roles (see the mean ranks and Kruskal-Wallis test results in Table 3). Given the results in Table 3, paired comparisons were then conducted using Mann-Whitney U tests to determine the specific differences between type pairs (e.g., between Team leads and Programmers).

Considering the practitioners assigned to the five roles, Table 3 shows that those who were assigned to fulfil Multiple roles typically used the highest level of individualistic language (e.g., I, me, my), and our Mann-Whitney results showed that the differences in use were statistically significant when this group of practitioners was compared with those occupying Admin ($p = 0.001$) and Project manager ($p = 0.027$) roles. A similar pattern of results is seen in Table 3 for positive language use (e.g., beautiful, relax, perfect); apart from the paired comparison with those occupying the Programmer role, our Mann-Whitney results confirmed that those that were assigned Multiple roles used significantly more positive language ($p < 0.01$) than others.

**Table 2**. Selected linguistic dimensions.

| Linguistic Category | Abbreviation (Abbrev.) | Examples | Reason for Inclusion |
|---|---|---|---|
| Pronouns | I | I, me, mine, my | Elevated use of first person plural pronouns (we) is evident during shared situations, whereas, relatively high use of self references (I) has been linked to individualistic attitudes (Pennebaker and Lay, 2002). Use of the second person pronoun (you) signals the degree to which members rely on other team members (Pennebaker et al., 2003). |
| | we | we, us, our, we've, lets | |
| | you | you, your, you'll, you've | |
| Cognitive language | cogmech | think, consider, determined, idea should, prefer, definitely, always, extremely, certain | Software teams were previously found to be most successful when many group members were highly cognitive and were natural solution providers (Andre et al., 2011). These traits are also linked to effective task analysis and brainstorming capabilities. |
| Work and achievement related language | work | feedback, goal, program, delegate, duty, meeting | Individuals most concerned with task completion and achievement are said to reflect these traits during their communication. Such individuals are most concerned with task success, contributing and initiating ideas and knowledge towards task completion (Benne and Sheats, 1948). |
| | achieve | accomplish, attain, resolve, finalize, solve | |
| Leisure, social and positive language | leisure | movie, entertain, party | Individuals that are personal and social in nature are said to communicate positive emotion and social words and this trait is said to contribute towards an optimistic group climate (Benne and Sheats, 1948). Leisure related language is also an indicator of a team-friendly atmosphere. |
| | social | give, love, explain, friend | |
| | posemo | beautiful, perfect, glad | |
| Negative language | negemo | afraid, crap, hate, dislike, terrified, suck, annoyed | Negative emotion affects team cohesiveness and group climate. This form of language shows discontent and resentment (Goldberg, 1981). |
| | anger | | |
| Past, present and future tenses | past | went, worked, accepted | Reflective communication is said to be evaluative, linking previous communications and adjusting previous viewpoints, and those focused on the present and future also communicate accordingly (Zhu, 1996). |
| | present | begin, does, try, completes | |
| | future | might, will, gonna, next | |
| Question mark | qmark | ? | Questions are said to start the knowledge construction process and help those responding to confirm their own understanding (Batson et al., 2002). |
| Word count | wc | - | People having more ideas and suggestions to convey generally communicate with longer messages (Gonzales et al., 2010). |

Table 3 further shows that Team leads made the most use of collective language (e.g., we, our, us), and our paired Mann-Whitney comparisons revealed statistically significant differences in this form of language use between this group of practitioners and those occupying the other roles: Multiple ($p = 0.002$), Admin ($p = 0.005$), Project manager ($p = 0.048$), and Programmer ($p = 0.000$), respectively. Similar patterns of results were obtained for reliance type language (e.g., you, your, you're), work and achievement language (e.g., feedback, goal, delegate), leisure language (e.g., club, movie, party), social language (e.g., give, buddy, love), future tense words, question mark use and message word count.

Table 3 shows that Project managers used the most cognitive language (e.g., think, believe, consider), negative emotion (e.g., afraid, hate, dislike) and past tense words. Our paired Mann-Whitney comparisons revealed statistically significant differences for cognitive language use when this group of practitioners was compared with those occupying Multiple ($p = 0.024$) and Programmer ($p = 0.003$) roles. Those occupying the Admin role used most present tense words, with statistically significant differences evident when these members were compared with those occupying Multiple ($p = 0.009$), Team lead ($p = 0.011$), and Programmer ($p = 0.000$) roles, respectively.

We then examined the roles enacted by those working in four of the teams selected in Table 1 (P1: user experience, P3: documentation, P7: code, and P8: project management) to investigate how those working on different forms of software tasks self-organize. We again conducted Kruskal-Wallis tests to check for differences in the 16 linguistic dimensions (shown in Table 2) between those occupying the five roles (Multiple, Team lead, Admin, Project manager and Programmer) for each of the four teams. These tests revealed that there were statistically significant differences ($p < 0.05$) in language usage among practitioners occupying the five roles for six of the linguistic dimensions for P1, three for P3, five for P7, and eight linguistic dimensions for P8. Due to space limitations we present these results in matrix form in Table 4, showing the role that dominated each of the linguistic dimensions for the four teams. Table 4 shows that a similar pattern of results is maintained (as seen in Table 3) for individualistic and positive language use, work word use and word count, and anger language use by those occupying Multiple, Team lead and Programmer roles, respectively. However, collective and negative language use was highest among four different roles for the four teams. A similar pattern of results is observed for the use of reliance, cognitive, achievement and leisure language, past tense, present tense, future tense and questions among the four teams, which varied among three groups of practitioners. Table 4 further shows that social language use was highest among Team leads and Admins.

**Table 3**. Mean ranks and Kruskal-Wallis test results.

| Linguistic Category | Abbrev. | Mean Rank | | | | | Kruskal-Wallis Test (*p*-value) |
|---|---|---|---|---|---|---|---|
| | | Multiple | Team lead | Admin | Project manager | Programmer | |
| Pronouns | I | **2991.03** | 2777.09 | 2361.71 | 2694.85 | 2810.56 | 0.008 |
| | we | 2622.86 | **2935.21** | 2537.20 | 2794.53 | 2703.34 | 0.000 |
| | you | 2674.87 | **2915.50** | 2613.93 | 2710.63 | 2723.35 | 0.000 |
| Cognitive language | cogmech | 2620.47 | 2891.15 | 2697.36 | **2960.21** | 2706.05 | 0.000 |
| Work and achievement related language | work | 2451.69 | **2932.13** | 2821.04 | 2716.94 | 2717.61 | 0.000 |
| | achieve | 2416.27 | **2942.57** | 2893.45 | 2656.81 | 2718.64 | 0.000 |
| Leisure, social and positive language | leisure | 2808.21 | **2906.31** | 2761.47 | 2791.67 | 2706.35 | 0.000 |
| | social | 2423.11 | **3022.93** | 2567.07 | 2809.93 | 2651.25 | 0.000 |
| | posemo | **3111.42** | 2652.12 | 2555.61 | 2429.64 | 2921.12 | 0.000 |
| Negative language | negemo | 2748.43 | 2790.96 | 2808.55 | **2856.85** | 2778.98 | 0.777 |
| | anger | 2759.90 | 2781.34 | 2725.04 | 2750.46 | **2802.29** | 0.278 |
| Past, present and future tenses | past | 2733.15 | 2809.96 | 2387.70 | **2845.99** | 2780.84 | 0.064 |
| | present | 2770.25 | 2886.07 | **3306.91** | 3063.61 | 2668.32 | 0.000 |
| | future | 2550.38 | **2898.23** | 2281.82 | 2733.33 | 2749.21 | 0.000 |
| Question mark | qmark | 2812.70 | **2851.67** | 2663.76 | 2769.12 | 2749.90 | 0.015 |
| Word count | wc | 2691.75 | **2982.79** | 2140.84 | 2626.31 | 2701.80 | 0.000 |

## 6. DISCUSSION, IMPLICATIONS AND FUTURE WORK

*RQ1. How are roles enacted during distributed agile software development?* Beyond their formally assigned roles, Jazz practitioners enacted various team-based roles to facilitate self-organisation during distributed software development. For the teams studied here, those that were assigned multiple roles enacted inter-personal team roles during project execution, while those that were formally assigned to administrative roles promoted task urgency. Project managers were expected to enact coordination and planning roles; however, our findings revealed that these members also performed more cognitive and insightful roles. On the other hand, team leaders enacted task-based and social roles and were integrally involved in task planning. Finally, programmers showed the most frustration during their teams developments (see details below).

*RQ2. Are the specific attitudes and competencies enacted related to practitioners' actual role assignment?* Jazz practitioners occupying various roles demonstrated a diverse range of attitudes and competencies during their projects. These differences were somewhat aligned to the types of features these practitioners were working on. Based on our analysis, and taking into account recent literature, it is our contention that the variances in attitudes and competencies observed among those occupying different roles are likely to have provided a balancing effect (Hoda et al., 2010), and may have therefore contributed to the Jazz teams' effective self-organization.

We noted that those in Jazz that were assigned to multiple roles were most individualistic; and these individuals were largely responsible for promoting a positive team climate. While evidence of individualistic behaviour denotes that these members were self-focused, and this trait is generally negative for teamwork

(Pennebaker and Lay, 2002), positive language use also promotes optimism and a team-friendly atmosphere (Benne and Sheats, 1948). Those who occupied the Admin role were particularly work and achievement focused; perhaps too much so, as these individuals were fixated with the specific issues under consideration. Involvement in configuration and integration of artefacts likely demands such an outlook, as a lack of focus can result in issues arising during integration that may lead to delays, or worse, project failures.

**Table 4**. Most pronounced language usage among roles for P1, P3, P7 and P8.

| Linguistic Category | Abbrev. | Project ID | | | |
|---|---|---|---|---|---|
| | | P1 | P3 | P7 | P8 |
| Pronouns | I | PM | Mul | Mul | Mul |
| | we | Admin | PM | Pgmr | TL |
| | you | TL | Admin | Admin | Mul |
| Cognitive language | cogmech | PM | TL | PM | Pgmr |
| Work and achievement related language | work | TL | TL | PM | TL |
| | achieve | Admin | TL | PM | TL |
| Leisure, social and positive language | leisure | Admin | TL | PM | TL |
| | social | TL | Admin | Admin | TL |
| | posemo | Mul | Mul | Admin | Mul |
| Negative language | negemo | Pgmr | Mul | PM | Admin |
| | anger | Pgmr | Pgmr | Pgmr | Mul |
| Past, present and future tenses | past | TL | Pgmr | PM | TL |
| | present | Admin | Mul | PM | Admin |
| | future | PM | TL | Pgmr | TL |
| Question mark | qmark | Pgmr | TL | PM | Pgmr |
| Word count | wc | TL | TL | PM | TL |

KEYS:- Mul = Multiple, TL = Team lead, PM = Project manager, Pgmr = Programmer

IBM Jazz project managers were highly cognitive and insightful during their projects, while programmers exhibited more cynical attitudes. Aligned with their actual role, project managers are required to support their teams with cognitive and insightful competencies in order to promote team confidence and effective project governance, and cynical attitudes exhibited by programmers may be linked to the significant mental challenges involved in coding. Although unconstructive attitudes such as these have a negative effect on team work (Goldberg, 1981), the positive, social and collective competencies exhibited by those occupying Admin, Team lead and Multiple roles could mitigate the negative effects of these undesirable attitudes.

IBM Jazz team leaders exhibited the greatest number of distinct competencies. Jazz team leaders maintained their teams' work and achievement focus, they contributed significantly to their teams' social climate, they were evaluative and reflective, and they were actively involved in planning for future tasks and activities. These members also had the most to say whenever they engaged during their projects. Given team leaders' actual responsibilities (leading, planning and integration), these skills were required of them for their projects to succeed. For instance, work and achievement focus promote team urgency, social climate is necessary for motivating others, and reflection and future planning is necessary to avoid repeating previous mistakes and for identifying then reducing likely future issues.

Our findings have implications for software development, and particularly for agile distributed teams. Our evidence here shows that intra-personal, inter-personal and organizational skills are required of all distributed agile software practitioners, but may be particularly necessary for those occupying leadership roles. The absence of these competencies may hamper project performance. In our future work, we plan to examine the way these practitioners' enacted roles, competencies and behaviours evolve over their projects from iteration to iteration, and to complement our psycholinguistic analysis with contextual (thematic) examinations. We encourage future research to conduct similar studies considering other distributed teams.

## 7. LIMITATIONS

The LIWC language constructs used to measure practitioners' attitudes and competencies have been used previously to study this subject, and were assessed for validity and reliability (Mairesse et al., 2007). However, the adequacy of these constructs (and the way the dimensions were combined) in the specific context of software development warrants further investigation. We reached data saturation (Glaser and Strauss, 1967) after analyzing artefacts from our third team, and our SNA results also confirmed that our sample was representative of those in the repository as we noted relative homogeneity in In-Degree measures across the ten projects. However, work processes and work culture at IBM are specific to that organization and may not be representative of organization dynamics elsewhere.

## ACKNOWLEDGEMENTS

S. Licorish is supported by an AUT VC Doctoral Scholarship. We thank IBM for granting us access to the Jazz repository. IBM, the IBM logo, ibm.com, and Rational are trademarks or registered trademarks of International Business Machines Corporation in the United States, other countries, or both.